\documentclass[journal,draftclsnofoot,11pt,onecolumn]{IEEEtran}
\IEEEoverridecommandlockouts
\usepackage{amsthm}
\usepackage{tabularx}
\usepackage{tabulary}
\usepackage{arydshln}
\usepackage{multirow}
\usepackage{calc}
\usepackage{blkarray}
\usepackage{url}
\usepackage{subcaption}
\usepackage[left=1.25in, right=1.25in, top=1in, bottom=1in]{geometry}
\usepackage{pgfplots}
\usepackage{siunitx}
\pgfplotsset{compat=1.12}
\usepackage{tikz}
\usepackage{verbatim}
\usetikzlibrary{shapes,arrows}
\usepackage{multicol,tabularx}
\usepackage{hhline}
\usepackage{enumitem}

\usetikzlibrary{positioning}
\usetikzlibrary{calc}
\usepackage{url}

\theoremstyle{definition}

\newtheorem{example}{Example}

\usepackage{graphicx,cite}
\usepackage{dblfloatfix}

\usepackage{blindtext, graphicx, amsfonts,
	amssymb,multirow,epstopdf}
\usepackage{dsfont}

\usepackage{mathtools}

\ifCLASSINFOpdf

\else

\fi

\newcommand{\calA}{\mathcal{A}}

\newcommand{\bfb}{\mathbf{b}}

\newcommand{\bfx}{\mathbf{x}}

\newcommand{\bfu}{\mathbf{u}}
\newcommand{\bfc}{\mathbf{c}}

\newcommand{\bfG}{\mathbf{G}}
\newcommand{\bfV}{\mathbf{V}}

\newcommand{\bfI}{\mathbf{I}}
\newcommand{\bfP}{\mathbf{P}}
\newcommand{\bfC}{\mathbf{C}}

\newcommand{\bfA}{\mathbf{A}}
\newcommand{\bfB}{\mathbf{B}}

\newcommand{\bfW}{\mathbf{W}}

\newcommand{\bfZ}{\mathbf{Z}}

\tikzset{pics/.cd,
  disc/.style={
    code={
      \fill [white] ellipse [x radius=2, y radius=2/3];
      \path [left color=black!50, right color=black!50, middle color=black!25]
        (-2+.05,-1.1) arc (180:360:2-.05 and 2/3-.05*2/3) -- cycle;
      \path [top color=black!25, bottom color=white]
        (0,.05*2/3) ellipse [x radius=2-.05, y radius=2/3-.05*2/3];
      \path [left color=black!25, right color=black!25, middle color=white]
        (-2,0) -- (-2,-1) arc (180:360:2 and 2/3) -- (2,0) arc (360:180:2 and 2/3);
      \foreach \r in {225,315}
        \foreach \i [evaluate={\s=30;}] in {0,2,...,30}
          \fill [black, fill opacity=1/50]
            (0,0) -- (\r+\s-\i:2 and 2/3) -- ++(0,-1)
            arc (\r+\s-\i:\r-\s+\i:2 and 2/3) -- ++(0,1) -- cycle;
      \foreach \r in {45,135}
        \foreach \i [evaluate={\s=30;}] in {0,2,...,30}
          \fill [black, fill opacity=1/50]
            (0,0) -- (\r+\s-\i:2 and 2/3)
            arc (\r+\s-\i:\r-\s+\i:2 and 2/3)  -- cycle;
    }
  },
  disc bottom/.style={
    code={
      \foreach \i in {0,2,...,30}
        \fill [black, fill opacity=1/60] (0,-1.1) ellipse [x radius=2+\i/40, y radius=2/3+\i/60];
      \path pic {disc};
    }
  }
}

\makeatletter
\tikzset{
    kosblock filldraw/.style n args={5}{
        draw=#4!#5, fill=#1!#2,#3, line width=3pt},
    kosblock rect/.style={
        block filldraw, rectangle},
    kosblock/.style={
        block rect, minimum height=0.8cm, minimum width=6em},
    from/.style args={#1 to #2}{
        above right={0cm of #1},
        /utils/exec=\pgfpointdiff
            {\tikz@scan@one@point\pgfutil@firstofone(#1)\relax}
            {\tikz@scan@one@point\pgfutil@firstofone(#2)\relax},
        minimum width/.expanded=\the\pgf@x,
        minimum height/.expanded=\the\pgf@y}
}
\makeatother

\tikzset{%
  block/.style    = {draw, thick, rectangle, minimum height = 1.5em,
    minimum width = 3em},
  recblock/.style    = {draw, thick, rectangle, minimum height = 1.7em,
    minimum width = 6em},
     rblock/.style    = {draw, thick, rectangle, minimum height = 1.7em,
    minimum width = 6.13em},
    reblock/.style    = {draw, thick, rectangle, minimum height = 1.7em,
    minimum width = 2em},
  sum/.style      = {draw, circle, node distance = 0.7cm}, 
  input/.style    = {coordinate}, 
  output/.style   = {coordinate} 
}

\tikzset{%
  liblock/.style    = {draw, thick, rectangle, minimum height = 3em,
    minimum width = 9.5em},
  lirecblock/.style    = {draw, thick, rectangle, minimum height = 1.7em,
    minimum width = 2em},
  lireblock/.style    = {draw, thick, rectangle, minimum height = 1.7em,
    minimum width = 4.5em},
  sum/.style      = {draw, circle, node distance = 1cm}, 
  input/.style    = {coordinate}, 
  output/.style   = {coordinate} 
}

\tikzset{%
block2/.style   = {draw, thick, rectangle, minimum height = 1.7em, minimum width = 10 em},
block1/.style   = {draw, thick, rectangle, minimum height = 1.7em, minimum width = 6 em}
}

\tikzset{
  custom dash/.style={dash pattern=on 10pt off 4pt},
}

\definecolor{koscolor1}{rgb}{0.00000,0.44700,0.74100}%
\definecolor{kosbluefilecolor}{rgb}{0,0,1}
\definecolor{kosgreenfilecolor}{rgb}{0.098, 0.827, 0.090}

\definecolor{mycolor1}{rgb}{0.00000,0.44700,0.74100}%
\definecolor{mycolor2}{rgb}{0.85000,0.32500,0.09800}%
\definecolor{mycolor3}{rgb}{0.92900,0.69400,0.12500}%
\definecolor{mycolor4}{rgb}{0.49400,0.18400,0.55600}%
\definecolor{mycolor5}{rgb}{0.46600,0.67400,0.18800}%
\definecolor{mycolor6}{rgb}{0.30100,0.74500,0.93300}%

\begin{document}
	\title{Straggler-resistant distributed matrix computation via coding theory}
\author{\IEEEauthorblockN{Aditya Ramamoorthy, Anindya Bijoy Das and Li Tang}\\
		\IEEEauthorblockA{Department of Electrical and Computer Engineering,\\
		Iowa State University,\\
		Ames, IA 50010\\
		Email: \{adityar,abd149,litang\}@iastate.edu}
}

\maketitle

\section{Introduction}
The current BigData era routinely requires the processing of large scale data on massive distributed computing clusters. In these applications, datasets are often so large that they cannot be housed in the memory and/or the disk of any one computer. Thus, the data and the processing is typically distributed across multiple nodes. Distributed computation is thus a necessity rather than a luxury. The widespread usage of such clusters presents several opportunities and advantages over traditional computing paradigms. However, it also presents newer challenges where coding-theoretic ideas have recently had a significant impact. Large scale clusters (which can be heterogeneous in nature) suffer from the problem of stragglers which refer to slow or failed worker nodes in the system. Thus, the overall speed of a computation is typically dominated by the slowest node in the absence of a sophisticated assignment of tasks to the worker nodes.

These issues are a potential bottleneck in several important and basic problems such as (but not limited to) the training of large scale models in machine learning. Operations such as matrix-vector multiplication and matrix-matrix multiplication (henceforth referred to as matrix computations) play a significant role in several parts of the machine learning pipeline \cite{goodfellow2016deep} ({\it cf.} Section \ref{sec:apps}). In this survey article, we overview recent developments in the field of coding for straggler-resilient distributed matrix computations.

\begin{figure}[t]
\centering
\resizebox{0.5\textwidth}{!}{
\begin{tikzpicture}[auto, thick, node distance=2cm, >=triangle 45]

\draw
	node at (0,0)[right=-3mm]{}
	node [reblock] (blk1){$W0$}
    node [reblock, right = 3cm of blk1] (blk2) {$W1$}
    node [reblock, right = 3cm of blk2] (blk3) {$W2$}
    node [rblock, minimum width = 3cm,below = 0.6 cm of blk1] (blk11) {$\bfA_{1}^T \, \bfx$}
    node [rblock, minimum width = 3cm,below = 0.0005 cm of blk11] (blk12) {$ \left( \bfA_{2}\, + \bfA_{3}\, \right)^T \bfx$}
    node [rblock, minimum width = 3cm,below = 0.6 cm of blk2] (blk21) {$\bfA_{2}^T\, \bfx$}
    node [rblock, minimum width = 3cm,below = 0.0005 cm of blk21] (blk22) {$\left( \bfA_{3}\, + \bfA_{1}\, \right)^T \bfx$}
    node [rblock, minimum width = 3cm,below = 0.6 cm of blk3] (blk31) {$\bfA_{3}^T \, \bfx$}
    node [rblock, minimum width = 3cm,below = 0.0005 cm of blk31] (blk32) {$\left( \bfA_{1}\, + \bfA_{2}\, \right)^T \bfx$}
    ;

\draw[->](blk1) -- node{} (blk11);
\draw[->](blk2) -- node{} (blk21);
\draw[->](blk3) -- node{} (blk31);

\end{tikzpicture}
}

\caption{\label{fig:intro_straggler} {\footnotesize Matrix $\bfA$ is split into three equal-sized block columns. Each node is responsible for computing submatrix-vector products, sequentially from top to bottom. Note that $\bfA^T \bfx$ can be decoded even if one node fails.}}
\vspace{-0.2in}
\end{figure}
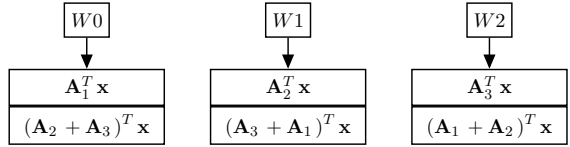

The conventional approach for tackling stragglers in distributed computation has been to run multiple copies of tasks on various machines \cite{ananthanarayanan2013effective}, with the hope that at least one copy finishes on time. However, coded computation offers significant benefits for specific classes of problems. We illustrate this by means of a matrix-vector multiplication example in Fig.\ref{fig:intro_straggler}. Consider the scenario where a user wants to compute $\bfA^T \bfx$ where $\bfA$ is a $t \times r$ matrix and $\bfx$ is a $t \times 1$ vector; both $t$ and $r$ are assumed to be large. The size of $\bfA$ precludes the possibility that the computation can take place on a single node. Accordingly, matrix $\bfA$ is block-column decomposed as $\bfA = [\bfA_1 ~\bfA_2~\bfA_3]$ where each $\bfA_i$ is of the same size. Each worker node is given the responsibility of computing two submatrix-vector products so that the computational load on each worker is $2/3$-rd of the original. We note here that the master node that creates the encoded matrices, e.g., $(\bfA_2 + \bfA_3)$ only needs to perform additions (and more generally scalar multiplications). The computationally intensive task of computing inner products of the rows of the encoded matrices with $\bfx$ is performed by the worker nodes. It can be observed that even if one worker is a complete straggler, i.e., it fails, there is enough information for a master node to compute the final result. This does however, require the master node to solve a linear system of equations to decode the final result. A similar approach (with additional subtleties) can be used to arrive at a corresponding illustrative example for matrix-matrix multiplication.

We note here that straggler mitigation using coding techniques has also been considered in a different body of work that broadly deals with reducing file access delays when retrieving data from cloud storage systems \cite{shahLR13,JoshiSW17,LiRS18,WangJW19}. Much of this work deals with understanding tradeoffs between file access latency and the redundancy introduced by the coding method under different service time models for the servers within the cloud. Coded systems in turn introduce interesting challenges in the queuing delay analysis of these systems. In this survey article, we will focus on the basic techniques needed for coded distributed matrix computation.

\section{Applications of matrix computations within distributed machine learning}
\label{sec:apps}


Computing high-dimensional linear transforms is an important component of dimensionality reduction techniques such as principal component analysis (PCA) and linear discriminant analysis (LDA) \cite{bishopML}. Large scale linear regression and filtering are also canonical examples of problems where linear transformations play a key role. They are also key components of training deep neural networks \cite{goodfellow2016deep} and using them for classification as we explain in detail below.

Every layer of a fully-connected deep neural network (see Fig. \ref{NN}) requires matrix-matrix multiplications in both forward and backward propagation. 
Suppose that the training data can be represented as a matrix $\bfP_0$ of size $f \times m$, where $f$ is the number of features and $m$ is the number of samples. In forward propagation, in any layer $i$ the input $\bfP_{i-1}$ is multiplied by the weight matrix $\bfW_i$ and the bias term $\bfb_{i}$ is added. Following this, it is passed through a non-linear function, $g_i(\cdot)$ to obtain $\bfP_i$ (the input of the next layer), i.e.,
\begin{align*}
\bfZ_i = \bfW_i \, \bfP_{i-1} \, + \,  \bfb_i \, \mathbf{1}^T  \; \; \; \; \; \textrm{and} \; \; \; \; \;  \bfP_i \; = \; g_i \left( \bfZ_i \right).
\end{align*} 
We note here that if $\bfW_i$ is a large matrix, then we have a large scale matrix-matrix multiplication problem that needs to be solved in this step.

\tikzset{%
  every neuron/.style={
    circle,
    draw,
    minimum size=0.6cm
  },
  neuron missing/.style={
    draw=none,
    scale=4,
    text height=0.333cm,
    execute at begin node=\color{black}$\vdots$
  },
}

\begin{figure}[t]
\centering
\resizebox{0.75\linewidth}{!}{
\begin{tikzpicture}[x=1.5cm, y=1.5cm, >=stealth]

\foreach \m/\l [count=\y] in {1,2,missing,3}
  \node [every neuron/.try, neuron \m/.try] (input-\m) at (0,2.5-\y) {};

\foreach \m [count=\y] in {1,2,3,4,missing,5}
  \node [every neuron/.try, neuron \m/.try ] (hidden1-\m) at (2,3.5-\y*1) {};

\foreach \m [count=\y] in {1,2,3,missing,4}
  \node [every neuron/.try, neuron \m/.try ] (hidden2-\m) at (4,3-\y*1) {};

  \foreach \m [count=\y] in {1,2,3,4,missing,5}
  \node [every neuron/.try, neuron \m/.try ] (hidden3-\m) at (6,3.5-\y*1) {};

\foreach \m [count=\y] in {1,missing,2}
  \node [every neuron/.try, neuron \m/.try ] (output-\m) at (8,2-\y) {};

\foreach \l [count=\i] in {1,2,f}
  \draw [<-] (input-\i) -- ++(-1,0)
    node [above, midway] {$X_\l$};

%

\foreach \l [count=\i] in {1,q}
  \draw [->] (output-\i) -- ++(1,0)
    node [above, midway] {$Y_\l$};

\foreach \i in {1,...,3}
  \foreach \j in {1,...,5}
    \draw [->] (input-\i) -- (hidden1-\j);

\foreach \i in {1,...,5}
  \foreach \j in {1,...,4}
    \draw [->] (hidden1-\i) -- (hidden2-\j);

\foreach \i in {1,...,4}
  \foreach \j in {1,...,5}
    \draw [->] (hidden2-\i) -- (hidden3-\j);

\foreach \i in {1,...,5}
  \foreach \j in {1,...,2}
    \draw [->] (hidden3-\i) -- (output-\j);

\draw [decorate,decoration={brace,amplitude=6pt},xshift=-4pt,yshift=0pt]
(1.5,2.8) -- (6.7,2.8) node [black,midway,xshift=-0.6cm]{};
\node at (4.1,3.2) {Hidden Units};
\draw [decorate,decoration={brace,amplitude=6pt},xshift=-4pt,yshift=0pt]
(1.5,2.8) -- (6.7,2.8) node [black,midway,xshift=-0.6cm]{};
\node at (4.1,3.2) {Hidden Units};

\draw [decorate,decoration={brace,amplitude=6pt},xshift=-4pt,yshift=0pt]
(7.3,2.8) -- (8.7,2.8) node [black,midway,xshift=-0.6cm]{};
\node at (7.9,3.2) {Output};
\draw [decorate,decoration={brace,amplitude=6pt},xshift=-4pt,yshift=0pt]
(7.3,2.8) -- (8.7,2.8) node [black,midway,xshift=-0.6cm]{};
\node at (7.9,3.2) {Output};

\draw [decorate,decoration={brace,amplitude=6pt},xshift=-4pt,yshift=0pt]
(-0.5,2.8) -- (0.9,2.8) node [black,midway,xshift=-0.6cm]{};
\node at (0.1,3.2) {Input};
\draw [decorate,decoration={brace,amplitude=6pt},xshift=-4pt,yshift=0pt]
(-0.5,2.8) -- (0.9,2.8) node [black,midway,xshift=-0.6cm]{};
\node at (0.1,3.2) {Input};

\end{tikzpicture}
}
\caption{\footnotesize A fully connected neural network with three hidden layers where an input vector has a size $f$ (number of features), and can be classified into one of $q$ classes.}
\label{NN}
\vspace{-0.2in}
\end{figure}
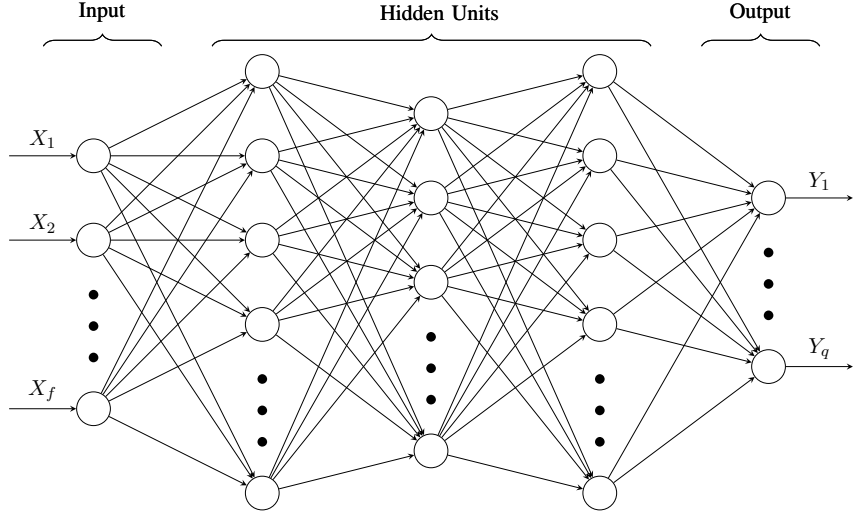

Similar issues also arise in the backpropagation step, where the weight matrices and bias vectors are adjusted. We typically use a variant of gradient descent to obtain the weight matrix $\bfW^j_i$ at iteration $j$ in layer $i$ using an appropriate learning rate $\alpha$. Now if $d \bfZ^j_i, d \bfW^j_i$ and $d \bfP^j_i$ indicate the gradients of the chosen loss function with respect to $\bfZ_i, \bfW_i$ and $\bfP_i$ respectively, then for any iteration $j$, we compute
\begin{align*}
d \bfZ^j_i \; &= \; g'_i \left( \bfZ^j_i \right) \odot d \bfP^j_i, \; \; \; \; \; \; \; \, \text{and} \; \; \; \;  d\bfW^j_i \; \; \; \, = \; \frac{1}{m}  \, d \bfZ^j_i \; \bfP_{i-1}^{j \; T} \, ; \\
\text{and update} \; \;  \bfW^j_i \; & = \; \bfW^{j-1}_i \, - \, \alpha \, d\bfW^j_i \; \; \; \; \; \; \text{and} \; \; \; \; \;  d\bfP^j_{i-1} \; = \;  \bfW^{j \; T}_i \; d \bfZ_{i}^{j} \, .
\end{align*}
The symbol $\odot$ above indicates the Hadamard product. This requires matrix-matrix multiplication in each layer as well. Furthermore, each of these steps is repeated over multiple iterations.

As a concrete example, consider AlexNet \cite{krizhevsky2012imagenet} which performs a 1000-way classification of the ImageNet dataset and provides a top-5 test error rate of under 15.3\%.
It has a training set of $1.2$ million images, $50,000$ validation images and a test set of $150,000$ images, each of which is a $224 \times 224 \times 3 \; ( = 150528)$ image. So, for training $\bfP_0$ has a size $\approx \num{1.5e05}$  by $\num{1.2e06}$. AlexNet consists of total eight layers, among which five are convolutional layers and the other three are fully connected layers.  Thus, this network has $43264$ and $4096$ neurons in the fifth and sixth layers,  so $\bfW_6$ has a size of $4096 \times 43264$. Thus, in the sixth layer of the forward propagation the network requires the product of two matrices of size $4096 \times 43264$ and $43264 \times (\num{1.2e06})$.

\section{Problem Formulation}
\label{sec:problem_form}
We present a formulation of the distributed matrix-matrix multiplication problem in this section. Note that matrix-vector multiplication is a special (though very important) case of matrix-matrix multiplication and the formulation carries over in this case in a natural manner. Consider a scenario where a master node has two large matrices  $\bfA \in \mathbb{R}^{t\times r}, \bfB \in \mathbb{R}^{t \times w}$ and wishes to compute $\bfA^T \bfB$ in a distributed fashion using $N$ worker nodes.

Each worker node is assigned a storage fraction for the coded columns of $\bfA$ (denoted $\gamma_A$) and $\bfB$ (denoted $\gamma_B$). The coded columns of $\bfA$ and $\bfB$ should be created by means of computationally inexpensive operations, e.g., scalar multiplications and additions. While the storage fraction constraint can be satisfied by potentially nonlinear coded solutions, our primary interest will be in linearly coded solutions where $\bfA$ and $\bfB$ are decomposed into block-matrices of size $p \times m$ and $p \times n$ respectively as shown below.
\begin{align}
\bfA &= \begin{bmatrix}
\bfA_{0,0} &\dots& \bfA_{0,m-1}\\
\vdots & \ddots & \vdots \\
\bfA_{p-1,0} & \dots & \bfA_{p-1,m-1}
\end{bmatrix}, \;\; \; \; \; \textrm{and} \; \; \; \; \;
\bfB = \begin{bmatrix}
\bfB_{0,0} &\dots& \bfB_{0,n-1}\\
\vdots & \ddots & \vdots \\
\bfB_{p-1,0} & \dots & \bfB_{p-1,n-1}
\end{bmatrix}, \label{eq:block_decomp_A_B}
\end{align}
so that the blocks in $\bfA$ and $\bfB$ are of size $\frac{t}{p} \times \frac{r}{m}$ and $\frac{t}{p} \times \frac{w}{n}$ respectively.
The master node generates certain linear combinations of the blocks in $\bfA$ and $\bfB$ and sends them to the worker nodes. The master node also requires each worker node to compute the product of some or all of their assigned matrices in a specified sequential order; we refer to this as the {\it responsibility} of the worker node. For instance, if a given worker node stores coded matrices $\tilde{\bfA}_0, \tilde{\bfA}_1$ and $\tilde{\bfB}_0, \tilde{\bfB}_1$ and is required to compute all four pairwise products, then the scheme specifies the order, e.g., $\tilde{\bfA}^T_0\tilde{\bfB}_0, \tilde{\bfA}^T_1\tilde{\bfB}_0, \tilde{\bfA}^T_0\tilde{\bfB}_1, \tilde{\bfA}^T_1\tilde{\bfB}_1$ or $\tilde{\bfA}^T_1\tilde{\bfB}_1, \tilde{\bfA}^T_1\tilde{\bfB}_0, \tilde{\bfA}^T_0\tilde{\bfB}_1, \tilde{\bfA}^T_0\tilde{\bfB}_0$ etc. The following two cases of block-partitioning $\bfA$ and $\bfB$ are of special interest.
\begin{itemize}
\item {\it Case 1 ($p=1$):} In this scenario, both $\bfA$ and $\bfB$ are decomposed into block columns, i.e., $\bfA = [\bfA_0 ~\bfA_1~\dots~\bfA_{m-1}]$ and $\bfB = [\bfB_0 ~\bfB_1~\dots~\bfB_{n-1}]$
so that recovering $\bfA^T \bfB$ is equivalent to recovering $\bfA_i^T \bfB_j$ for all pairs $i=0, \dots, m-1$, $j = 0, \dots, n-1$.
\item {\it Case 2 ($m=n=1$):} We set $\bfA^T = [\bfA^T_0 ~\bfA^T_1~\dots~\bfA^T_{p-1}]$ and $\bfB^T = [\bfB^T_0 ~\bfB^T_1~\dots~\bfB^T_{p-1}]$ so that $\bfA^T \bfB = \sum_{i=0}^{p-1} \bfA_i^T \bfB_i$.
\end{itemize}
The computational cost of computing $\bfA^T \bfB$ is $rw(2t -1)$ floating point operations (flops) which is approximately $\text{cost}(r,t,w) = 2rtw$ when $t$ is large. In the distributed setup under consideration, the computational load on each worker node is lesser than the original cost of computing $\bfA^T \bfB$ and the advantages of parallelism can therefore be leveraged.

We note some minor differences in the matrix-vector multiplication scenario at this point. Here, the master node wishes to compute $\bfA^T \bfx$, where $\bfx$ is a vector. As $\bfx$ is much smaller as compared to $\bfA$, we typically only impose the storage constraint for the worker nodes for the matrix $\bfA$ and assume that $\bfx$ is available to all of them. The case when $\bfx$ is further split into sub-vectors \cite{dutta2016short} will be treated along with the matrix-matrix multiplication case. 

\begin{example}
\label{eg:polycode}
Consider distributed matrix multiplication with $p=1$ and $m=n=2$. Furthermore, we define the matrix polynomials
\begin{align*}
\bfA(z) &= \bfA_0 + \bfA_1 z, \; \; \text{~and} \;\; \; \; \; \bfB(z) = \bfB_0 + \bfB_1 z^2 ;\\
\textrm{so that} \; \; \; \;  \; \bfA^T(z) \bfB(z) \; &= \; \bfA^T_0 \bfB_0 + \bfA^T_1\bfB_0 z + \bfA^T_0 \bfB_1 z^2 + \bfA^T_1 \bfB_1 z^3.
\end{align*}
Suppose that the master node evaluates $\bfA(z)$ and $\bfB(z)$ at distinct points $z_1, \dots, z_N$. It sends $\bfA(z_i)$ and $\bfB(z_i)$ to the $i$-th worker node, which is assigned the responsibility of computing  $\bfA^T(z_i)\bfB(z_i)$. 
It follows that as soon as {\it any} four out of the $N$ worker nodes return the results of their computation, the master node can perform polynomial interpolation to recover the $(k,l)$-th entry of each $\bfA_i^T\bfB_j$ for $0 \leq k < r/2$ and $0 \leq l < w/2$. Therefore, such a system is resilient to $N-4$ failures.

Note here that each worker node stores coded versions of $\bfA$ and $\bfB$ of size $t \times r/2$ and $t \times w/2$ respectively, i.e., $\gamma_A = \gamma_B = 1/2$. The computational load on each worker is $\text{cost}(r/2,t,w/2) =  \text{cost}(r,t,w)/4$, i.e., $1/4$-th of the original. Furthermore, each worker communicates a $r/2 \times w/2$ matrix to the master node.
\end{example}
\noindent On the other hand, splitting the matrices as in Case 2, yields a different tradeoff.
\begin{example}
\label{eg:entangled_poly_code}
Let $m=n=1$ and $p=2$, so that $ \bfA = \begin{bmatrix}
\bfA_0\\
\bfA_1
\end{bmatrix} \; \textrm{and} \;
\bfB = \begin{bmatrix}
\bfB_0\\
\bfB_1
\end{bmatrix}$ and consider the following matrix polynomials
\begin{align*}
\bfA(z) &= \bfA_0 z + \bfA_1 \;\; \; \; \; \textrm{and} \; \; \; \; \;
\bfB(z) = \bfB_0 + \bfB_1 z, \; \\
\textrm{so that} \; \; \; \;  \; \bfA^T(z) \bfB(z) &= \bfA_1^T \bfB_0 + (\bfA_0^T \bfB_0 + \bfA_1^T \bfB_1) z + \bfA_0^T \bfB_1 z^2.
\end{align*}
As before, the master node evaluates $\bfA(z)$ and $\bfB(z)$ at distinct points $z_1, \dots, z_N$ and sends the coded matrices to the worker nodes who calculate $\bfA^T(z_i)\bfB(z_i)$. In this case, as soon as any three workers complete their tasks, the master node can interpolate to recover $\bfA^T(z) \bfB(z)$ and obtain the desired result $(\bfA_0^T \bfB_0 + \bfA_1^T \bfB_1)$ as the coefficient of $z$. The other coefficients are interference terms. Thus, this system is resilient to $N-3$ stragglers and strictly improves on Example \ref{eg:polycode}, with the same storage fraction $\gamma_A = \gamma_B = 1/2$.

The dimensions of $\bfA(z)$ and $\bfB(z)$ are $t/2 \times r$ and $t/2 \times w$ so that the computational load on each worker is $\text{cost}(r,t/2,w) = \text{cost}(r,t,w)/2$, i.e., it is twice that of the workers in Example \ref{eg:polycode}. Moreover, each worker node communicates a $r \times w$ matrix to the master node, i.e., the communication load is four times that of Example \ref{eg:polycode}. 
\end{example}

\subsection{Metrics for evaluating coded computing solutions}
\label{sec:metrics_discussion}
Examples \ref{eg:polycode} and \ref{eg:entangled_poly_code} illustrate the core metrics by which coded computing solutions are evaluated. More formally, for given storage fractions $\gamma_A$ and $\gamma_B$ and the responsibilities of all the worker nodes, we evaluate a solution by a subset of the following metrics.
\begin{itemize}[wide, labelwidth=!, labelindent=0pt]
\item {\it Recovery threshold.} We say that a solution has recovery threshold $\tau$ if $\bfA^T \bfB$ can be decoded by the master node as long as {\it any} $\tau$ worker nodes return the results of their computation, e.g., the thresholds were four and three respectively in Examples \ref{eg:polycode} and \ref{eg:entangled_poly_code} above. This metric is most useful under the assumption that worker nodes are either working properly or in failure.
\item {\it Recovery threshold(II).} A more refined notion of recovery is required when we consider scenarios where worker nodes may be slow, but not complete failures. 
    For instance Fig. \ref{recthreshQ} shows an example where each worker node is assigned two matrix-vector products and operates sequentially from top to bottom. It can be verified by inspection that as long as any three matrix-vector products are obtained from the worker nodes in this manner, the master node has enough information to decode $\bfA^T \bfx$. For instance, Fig. \ref{recthreshQ} (left side) depicts a situation where $W2$ is failed and $W0$ is slow as compared to $W1$. The solution leverages the partial computations of $W0$ as well. We say that a solution has a recovery threshold(II) of $\tau'$ if the master node can decode the intended result if it receives the result of $\tau'$ computations from the worker nodes; these computations have to respect the sequential order within each worker node.


\begin{figure}[t]
\centering
\resizebox{0.86\linewidth}{!}{
\begin{tikzpicture}[auto, thick, node distance=2cm, >=triangle 45]

\draw

    node [sum, fill=blue!30,] (blk1) {$W0$}
    node [sum, fill=blue!30,right = 2.2cm of blk1] (blk2) {$W1$}
    node [sum, fill=blue!30,right = 2.2cm of blk2] (blk3) {$W2$}

    node [block, minimum width = 2.8cm, fill=green!30,below = 0.4 cm of blk1] (blk11) {$\bfA_{1}^T \bfx$}
    node [block, minimum width = 2.8cm,below = 0.0005 cm of blk11] (blk12) {$ \left( \bfA_{2} + \bfA_{3} \right)^T \bfx$}

    node [block, minimum width = 2.8cm,fill=green!30,below = 0.4 cm of blk2] (blk21) {$\bfA_{2}^T \bfx$}
    node [block, minimum width = 2.8cm,fill=green!30,below = 0.0005 cm of blk21] (blk22) {$\left( \bfA_{3} + \bfA_{1} \right)^T \bfx$}
	
    node [block, minimum width = 2.8cm,below = 0.4 cm of blk3] (blk31) {$ \bfA_3^T \bfx$}
    node [block, minimum width = 2.8cm,below = 0.0005 cm of blk31] (blk32) {$ \left( \bfA_1 + \bfA_{2} \right)^T \bfx$}

    node [sum, fill=blue!30,right = 4.2cm of blk3] (blk4) {$W0$}
    node [sum, fill=blue!30,right = 2.2cm of blk4] (blk5) {$W1$}
    node [sum, fill=blue!30,right = 2.2cm of blk5] (blk6) {$W2$}

    node [block, minimum width = 2.8cm,fill=green!30,below = 0.4 cm of blk4] (blk41) {$\bfA_{1}^T \bfx$}
    node [block, minimum width = 2.8cm,below = 0.0005 cm of blk41] (blk12) {$ \left( \bfA_{2} + \bfA_{3} \right)^T \bfx$}

    node [block, minimum width = 2.8cm,fill=green!30, below = 0.4 cm of blk5] (blk51) {$\bfA_{2}^T \bfx$}
    node [block, minimum width = 2.8cm,below = 0.0005 cm of blk51] (blk22) {$\left( \bfA_{3} + \bfA_{1} \right)^T \bfx$}
	
    node [block, minimum width = 2.8cm,fill=green!30,below = 0.4 cm of blk6] (blk61) {$ \bfA_3^T \bfx$}
    node [block, minimum width = 2.8cm,below = 0.0005 cm of blk61] (blk32) {$ \left( \bfA_{1} + \bfA_2 \right)^T \bfx$}
	
;
\draw[->](blk1) -- node{} (blk11);
\draw[->](blk2) -- node{} (blk21);
\draw[->](blk3) -- node{} (blk31);
\draw[->](blk4) -- node{} (blk41);
\draw[->](blk5) -- node{} (blk51);
\draw[->](blk6) -- node{} (blk61);

\end{tikzpicture}
}

\caption{\footnotesize The figure depicts two example scenarios, where the master node obtains the results of three completed tasks (respecting the sequential order) from the worker nodes. The scheme is such that the master node is guaranteed to recover $\bfA^T \bfx$ as long as any three tasks are completed.}
\label{recthreshQ}
\vspace{-0.2in}
\end{figure}
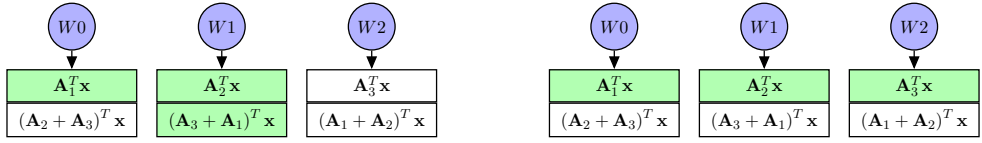 

\item {\it Computational load per worker node.} The complexity of determining $\bfA^T \bfB$ is $\text{cost}(r,t,w)$ flops. The computational load per worker is measured as a fraction of $\text{cost}(r,t,w)$, e.g., in Examples \ref{eg:polycode} and \ref{eg:entangled_poly_code}, the fractions are $1/4$ and $1/2$ respectively. We note here that if $\bfA$ and $\bfB$ are sparse then the computational load on the worker will depend on the number of non-zero entries in them. We discuss this point in more detail in Section \ref{sec:opps_future_work}.
\item {\it Communication load per worker node.} The communication load per worker measures the number of values that a worker node needs to send to the master node, normalized by $rw$.
\item {\it Decoding Complexity.} All linear schemes under consideration in this article require solving a system of linear equations to decode the result $\bfA^T \bfB$. The time-complexity of solving an arbitrary $\ell \times \ell$ system of equations grows as $\ell^3$. This is another metric that needs to be small enough for a scheme to be useful. For instance, in Example \ref{eg:polycode} the master node needs to solve a $4 \times 4$ system of equations, $rw/4$ times. Thus, the time-cost of decoding is roughly proportional to $rw$; there is no dependence on $t$. On the other hand the computation load on a worker does depend in a multiplicative manner on $t$. In scenarios where $t$ is large, it can be argued that the decoding cost is negligible compared to the worker computation. Nevertheless, we point out that this is a metric that needs to be taken into account. We note here that decoding in Examples \ref{eg:polycode} and \ref{eg:entangled_poly_code} corresponds to polynomial interpolation and is thus a ``structured" system of equations that can be typically solved much faster than Gaussian elimination.
\item {\it Numerical stability.} Solving linear equations to determine $\bfA^T \bfB$ naturally brings up the issue of numerical stability of the decoding. Specifically, if the system of equations is ill-conditioned, then the decoded result may suffer from significant numerical inaccuracies. Let $\bfP$ be a real-valued matrix and $\sigma_{\max}(\bfP)$ and $\sigma_{\min}(\bfP)$ denote its maximum and minimum singular values \cite{horn1990matrix}. We define its condition number
    \begin{align*}
    \text{cond}(\bfP) &= \frac{\sigma_{\max}(\bfP)}{\sigma_{\min}(\bfP)}.
    \end{align*}
    As a thumb-rule, if the system of equations has a condition number of $10^l$ it results in the loss of approximately $l$-bits of numerical precision. For any distributed scheme, we ideally want the worst case condition number over all possible recovery matrices to be as small as possible.
\end{itemize}

\section{Overview of Techniques}
\label{sec:overview_techniques}

The overarching idea in almost all of the works in this area is one of ``embedding" the matrix computation into the structure of an erasure code. Note that $(n,k)$ erasure codes \cite{lincostello} used in point-to-point communication have the property that one can decode the intended message for several erasure patterns, e.g., for maximum distance separable (MDS) codes as long as any $k$ coded symbols (out of $n$) are obtained, the receiver can decode the intended message. Most constructions of MDS codes are non-binary and decoding typically involves multiplications and divisions. There are also several constructions of binary, near-MDS codes, e.g.,  LDPC codes and fountain codes that allow recovery with high probability from any $k(1+\epsilon)$ symbols for small $\epsilon > 0$ when $n$ and $k$ are large. The decoding can be performed by simple add/subtract operations.

For instance, Example \ref{eg:polycode} demonstrates an embedding of matrix-matrix multiplication into the structure of a Reed-Solomon code. It can be observed that this embedding essentially requires that the $i$-th worker node computes the evaluation of polynomial $\bfA^T(z) \bfB(z)$ at $z_i$; this evaluation may or may not be received based on whether the $i$-th worker node is a straggler. In contrast, in the traditional communication scenario, the transmitter computes the evaluation and the channel uncertainty dictates whether or not the evaluation is received. Moreover, the decoding in Example \ref{eg:polycode} corresponds to polynomial interpolation which is precisely what Reed-Solomon decoding (from erasures) amounts to. Despite the similarities, we emphasize that in the matrix computation setup we operate within the real field $\mathbb{R}$, while traditional erasure coding almost exclusively considers operations over finite fields. As we will see, this introduces additional complications in the distributed computation scenario.

The original idea of using redundancy to protect against node failures in distributed matrix computation goes back to the work on ``algorithm-based fault tolerance"  from the 80's \cite{huangA84,jouA86}. However, more recent contributions have significantly improved on them. Ideas from polynomial evaluation and interpolation have played an important role in this area. We briefly recapitulate some of these ideas below.

\subsection{Primer on polynomials}
\label{sec:primer_poly}
Let $u(z) = \sum_{k=0}^d u_k z^k$ be a polynomial of degree $d$ with real coefficients. Let $u^{(j)}(z)$ denote the $j$-th derivative of $u(z)$. It can be verified that
\begin{align}
	u^{(j)}(z) = \sum_{k=0}^d u_k \binom{k}{j} j!~ z^{k-j}, \label{eq:real_poly_der}
\end{align}
where $\binom{k}{i} = 0$ if $k < i$.
Furthermore, note that we can also represent $u(z)$ by considering its Taylor series expansion around a point $\beta \in \mathbb{R}$, i.e.,
\begin{align}
	u(z) = \sum_{k=0}^d \frac{u^{(k)}(\beta)}{k!} (z - \beta)^k. \label{eq:real_Taylor_series}
\end{align}
It is well known that $u(z)$ has a zero of multiplicity $\ell$ at $\beta \in \mathbb{R}$ if and only if $u^{(i)}(\beta) = 0$ for $0 \leq i < \ell$ and $u^{(\ell)}(\beta) \neq 0$.

Another well known fact states that if we obtain $d+1$ evaluations of $u(z)$ at distinct points $z_1, \dots, z_{d+1}$ then we can interpolate to find the coefficients of $u(z)$. This follows from the fact that the Vandermonde matrix $\bfV$ with parameters $z_1, \dots, z_{d+1}$, i.e., $\bfV_{i,j} = z_j^{i}$ for $0 \leq i \leq d, 1 \leq j \leq d+1$ is nonsingular when $z_j, j = 1,\dots,d+1$ are distinct. An interesting generalization holds when we consider not only the evaluations of $u(z)$ but also its derivatives. We illustrate this by the following example.
%
\begin{example}
Let $d=2$, so that $u(z) = u_0 + u_1 z +  u_2 z^2$ and the first derivative $u^{(1)}(z) = u_1 + 2 u_2 z$. Suppose that we obtain $u(z_1), u^{(1)}(z_1)$ and $u(z_2)$, where $z_1 \neq z_2$. We claim that this suffices to recover $u(z)$. To see this assume otherwise, i.e., there exists $\tilde{u}(z) \neq u(z)$ such that $u(z_1) = \tilde{u}(z_1), u^{(1)}(z_1) = \tilde{u}^{(1)}(z_1)$ and $u(z_2)= \tilde{u}(z_2)$. This in turn implies that there exists a polynomial $a(z) = u(z) - \tilde{u}(z)$ such that $a(z_1)= a^{(1)}(z_1)=a(z_2)=0$. Now, we note that $a(z)$ is such that it has a zero of multiplicity 2 at $z_1$ and a zero of multiplicity $1$ at $z_2$. The fundamental theorem of algebra states that if a polynomial has more zeros (counting multiplicities) than its degree, then it has to be identically zero. Therefore, we can conclude that $a(z)$ is identically zero and we can recover $u(z)$ exactly. This can also be equivalently be seen by examining
\begin{align*}
\det \begin{bmatrix}
1 & 0 & 1\\
z_1 & 1 & z_2\\
z_1^2 & 2z_1 & z_2^2
\end{bmatrix} \; = \; (z_2^2 - 2z_1z_2) + (2z_1^2 - z_1^2) \; = \;  (z_1 - z_2)^2 > 0.
\end{align*}
\end{example}
In general, for a polynomial $u(z) = \sum_{k=0}^d u_k z^k$ of degree $d$, suppose that we obtain $u^{(\ell)}(z_i)$ for $\ell = 0, \dots, k_i-1$ (where $u^{(0)}(z) = u(z)$) for distinct points $z_i, i = 1, \dots, N$. In this case if $\sum_{i=1}^N k_i \geq d+1$, then we can recover $u(z)$ exactly \cite{sobczyk2002}.

Note that polynomial interpolation is equivalent to solving a Vandermonde system of equations. However, since this system of equations is structured, the complexity can be reduced.
Specifically, a degree-$d$ polynomial can be interpolated with time-complexity $O(d \log^2 d)$ \cite{Pan2013TR}.
%
%
%
\begin{table}[t]
\centering
\caption{\label{table1} {\small Worst-case condition Numbers for the different schemes. For the Vandermonde scheme, the parameters are spaced uniformly in $[-1,1]$. For the ``\cite{RamTV19} + Embedding" scheme, each worker node is assigned two matrix-vector products corresponding to the polynomial evaluation and its first derivative. The embedding matrix $\bfC$ corresponds to the matrix representation of $GF(3^3)$.}}
\begin{tabular}{c c c c c} 
\hline
Scenario & Vandermonde &  \cite{RamTV19}+Embedding & \cite{das2019random} (ones) &  \cite{das2019random} (random) \\ \hline
$N = 15, \tau = 13$ & $1.689 \times 10^6$  & $411$ & $910$ & $264.49$ \\ \hline
$N = 15, \tau = 12$ & $1.695 \times 10^6$  & $949$ & $1.066 \times 10^4$ & $1.111 \times 10^3$\\ \hline
$N = 30, \tau = 28$ & $2.293 \times 10^{13}$ & $-$ & $2868.32$ & $1374.59$ \\ \hline
\end{tabular}%
\vspace{-0.2in}
\end{table}%

\subsection{Distributed Matrix-Vector Multiplication}
\label{sec:distr_mat_vec}
In more recent times, the power of coding-theoretic methods for matrix-vector multiplication was first explored in the work of Lee et al. \cite{lee2018speeding}. 
In the notation of Section \ref{sec:problem_form} set $p=1$ and consider splitting $\bfA = [\bfA_0~\bfA_1~\dots~\bfA_{m-1}]$ into $m$ equal-sized block columns. Here $m$ is a parameter that is a design choice. The idea of \cite{lee2018speeding} is to pick the generator matrix of a $(N,m)$ MDS code denoted $\bfG = (g_{ij}) \in \mathbb{R}^{m \times N}$. The master node then computes
\begin{align*}
\tilde{\bfA}_l &= \sum_{i=0}^{m-1} g_{il} \bfA_i
\end{align*}
and distributes $\bfx$ and $\tilde{\bfA}_l$ to the $l$-th worker node for $l = 0, \dots, N-1$, which computes $\tilde{\bfA}^T_l \bfx$.  The master node wishes to decode $\bfA_j^T \bfx, j = 0, \dots m-1$. Suppose that worker nodes indexed by $i_0, \dots, i_{m-1}$ are the first $m$ nodes to return their results. Note that the master node has
\begin{align*}
\tilde{\bfA}^T_{i_l} \bfx &= \sum_{j=0}^{m-1} g_{j i_l} (\bfA^T_i \bfx), \text{~for $l=0, \dots, m-1$},
\end{align*}
which implies that it can solve a system of linear equations to determine the required result if the $m \times m$ submatrix of $\bfG$ indexed by columns $i_0, \dots, i_{m-1}$ is non-singular; the MDS property of $\bfG$ guarantees this. 
Typical choices of $\bfG$ include picking it as a Vandermonde matrix with distinct parameters $z_1, \dots, z_N$. 
In this case, each $\tilde{\bfA}_l$ is the evaluation of $\bfA(z) = \bfA_0 + \bfA_1 z + \dots + \bfA_{m-1} z^{m-1}$ at $z = z_l$. 
The recovery threshold is $m$, the computational and communication load of each worker node is $1/m$-th of the original and the decoding can be performed faster than Gaussian elimination ({\it cf.} Section \ref{sec:primer_poly}). However, numerical stability is a significant concern when $\bfG$ has the Vandermonde form. It is well-known from the numerical analysis literature \cite{pan_Vandermonde16} that the condition number of a $\ell \times \ell$ real Vandermonde matrix grows exponentially in $\ell$. Table \ref{table1}, Column 2, contains some illustrative figures. It shows that even for $N=30$ with a threshold $\tau = 28$, the condition number is too high to be useful in practice.

On the other hand choosing each entry of $\bfG$ i.i.d. at random from a continuous distribution also works with high probability and the computational load per worker is still $1/m$-th of the original. Numerical stability is better \cite{subramaniamHN19}; however, decoding the system of equations will typically take time which is cubic in the size of the system of equations.


One can also use the idea of using polynomial interpolation with multiplicities discussed in Section \ref{sec:primer_poly} above.  Let the $j$-th derivative of $\bfA(z)$ be defined as follows.
\begin{align*}
\bfA^{(j)}(z) = \sum_{k=0}^{m-1} \bfA_k \binom{k}{j} j!~ z^{k-j}. \label{eq:matrix_poly_der}
\end{align*}
Suppose that the storage fraction $\gamma_A = 2/m$. In this case, for the $i$-th worker node, the master node assigns the computation of first  $[\bfA(z_i)]^T \bfx $ and then $[\bfA^{(1)}(z_i)]^T \bfx$.
As soon as a worker node completes a task, it sends the result to the master node. The result of \cite{RamTV19} demonstrates that as long as the master node receives $m$ matrix-vector multiplication results, it can decode the intended result, i.e., its recovery threshold(II) is $m$. The computational and communication load of each worker is $2/m$-th of the original. The key advantage of this scheme is that it allows the master node to leverage partial computations performed by slow nodes. However, numerical stability continues to be a problem here.

\begin{figure}[t]
\centering
\resizebox{0.75\linewidth}{!}{
\begin{tikzpicture}[auto, thick, node distance=2cm, >=triangle 45]

\draw

    node [sum, fill=blue!30] (blk1) {$W0$}
    node [sum, fill=green!30,right = 5cm of blk1] (blk2) {$W1$}
    node [sum, fill=blue!30,below = 3.7cm of blk1] (blk3) {$W2$}
    node [sum, fill=green!30,right = 5cm of blk3] (blk4) {$W3$}

    node [block, fill=blue!30, minimum width = 15.5em, below = 0.4 cm of blk1] (blk11) {$\left( \bfA_0 + \bfA_3 + \bfA_6 \right) $}
    node [block, fill=blue!30,minimum width = 15.5em,below = 0.0005 cm of blk11] (blk12) {$\left( \bfA_1 + \bfA_4 + \bfA_7 \right) $}
    node [block, fill=blue!30,minimum width = 15.5em,below = 0.0005 cm of blk12] (blk13) {$\left( \bfA_2 + \bfA_5 + \bfA_8 \right) $}
    node [block, fill=orange!30,minimum width = 15.5em,below =  0.1 cm of blk13] (blk17) {$\bfx$}

    node [block, fill=green!30,minimum width = 13em,below = 0.4 cm of blk2] (blk21) {$\left( \bfA_0 + \bfA_4 + \bfA_8 \right) $}
    node [block, fill=green!30,minimum width = 13em,below = 0.0005 cm of blk21] (blk22) {$\left( \bfA_1 + \bfA_5 + \bfA_6 + \bfA_7 \right) $}
    node [block, fill=green!30,minimum width = 13em,below = 0.0005 cm of blk22] (blk23) {$\left( \bfA_2 + \bfA_3 + \bfA_4 + \bfA_7 + \bfA_8 \right) $}
	node [block, fill=orange!30,minimum width = 13em,below =  0.1 cm of blk23] (blk27) {$\bfx$}
	
    node [block, fill=blue!30,minimum width = 15.5em,below = 0.4 cm of blk3] (blk31) {$\left( \bfA_0 + \bfA_5 + \bfA_7 + \bfA_8 \right) $}
    node [block, fill=blue!30,minimum width = 15.5em,below = 0.0005 cm of blk31] (blk32) {$\left( \bfA_1 + \bfA_3 + \bfA_4 + \bfA_6 + \bfA_7 + \bfA_8 \right) $}
    node [block, fill=blue!30,minimum width = 15.5em,below = 0.0005 cm of blk32] (blk33) {$\left( \bfA_2 + \bfA_4 + \bfA_5 + \bfA_6 + \bfA_8 \right) $}
	node [block, fill=orange!30,minimum width = 15.5em, below =  0.1 cm of blk33] (blk37) {$\bfx$}
	
    node [block, fill=green!30,minimum width = 13em, below = 0.4 cm of blk4] (blk41) {$\left( \bfA_0 + \bfA_3 + \bfA_4 + \bfA_6 + \bfA_8 \right) $}
    node [block, fill=green!30,minimum width = 13em,below = 0.0005 cm of blk41] (blk42) {$\left( \bfA_1 + \bfA_4 + \bfA_5 + \bfA_6 \right) $}
    node [block, fill=green!30,minimum width = 13em,below = 0.0005 cm of blk42] (blk43) {$\left( \bfA_2 + \bfA_3 + \bfA_4 + \bfA_5 + \bfA_7 \right) $}
	node [block, fill=orange!30,minimum width = 13em, below =  0.1 cm of blk43] (blk47) {$\bfx$}
;
\draw[->](blk1) -- node{} (blk11);
\draw[->](blk2) -- node{} (blk21);
\draw[->](blk3) -- node{} (blk31);
\draw[->](blk4) -- node{} (blk41);

\end{tikzpicture}
}

\caption{\footnotesize The scheme corresponding to the approach of \cite{RamTV19} as described in Example \ref{eg:udm_eg}.}
\label{unidecex}
\vspace{-0.2in}
\end{figure}
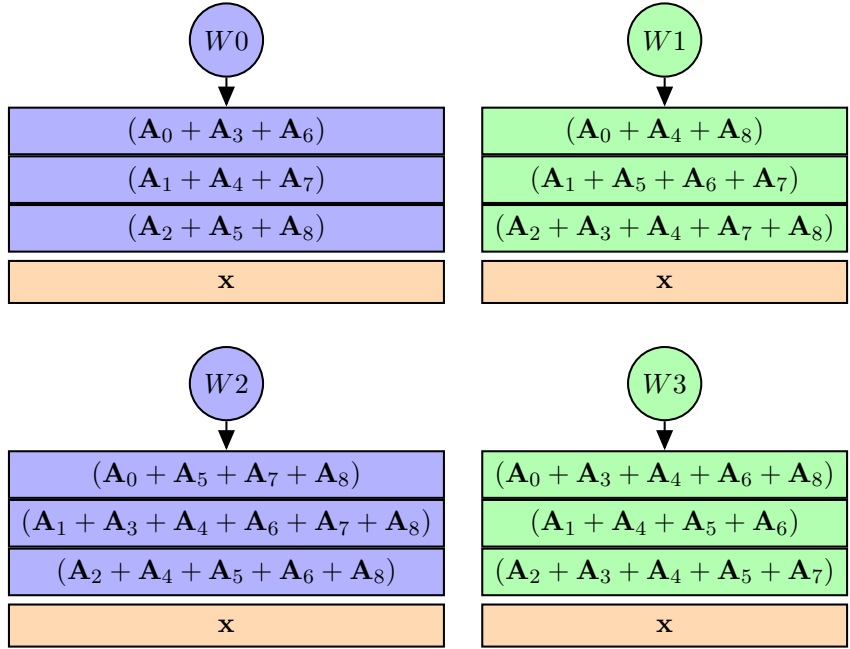

The numerical stability issue with both approaches discussed above can be addressed (to a certain extent) by a related idea that involves polynomials over finite fields. In particular, one can define polynomials over finite fields and their corresponding Hasse derivatives (resulting in so-called universally decodable matrices) and use an isomorphism between finite field elements and appropriate matrices to arrive at ``binary" schemes that have much better behaved condition number. We illustrate the basic idea by means of an example below and refer the reader to \cite{RamTV19} for the full details.
\begin{example}
\label{eg:udm_eg}
Let $u(z) = u_0 + u_1 z + u_2 z^2$ be a polynomial of degree-2. The discussion in Section \ref{sec:primer_poly} indicates that an associated $3 \times 3$ Vandermonde matrix is non-singular when the polynomial is evaluated at distinct points $z_1, z_2$ and $z_3$. It turns out that we can instead evaluate the polynomial at appropriately defined matrices instead and obtain schemes with useful properties. Let binary matrix $\bfC$ correspond to the matrix representation of the finite field $GF(3^3)$ (see \cite{RamTV19} and \cite{ward94matrix} for details), and consider powers of $\bfC$, i.e., $\bfC^\ell$ reduced modulo-2, as
\begin{align*}
\bfC = \begin{bmatrix}
0 & 0 & 1\\
1 & 0 & 1\\
0 & 1 & 0
\end{bmatrix} \; \; \; \textrm{and so, e.g.,} \;  \; \; \bfC^2 = \begin{bmatrix}
0 & 1 & 0\\
0 & 1 & 1\\
1 & 0 & 1
\end{bmatrix} \mod 2.
\end{align*}
Consider the $\bfG$ specified below (where each power of $\bfC$ is reduced modulo-2).
\begin{align*}
\bfG = \begin{bmatrix}
\bfI & \bfI & \bfI & \bfI\\
\bfI & \bfC & \bfC^2 & \bfC^3\\
\bfI & \bfC^2 & \bfC^4 & \bfC^6
\end{bmatrix}
\end{align*}
The work of \cite{RamTV19} shows, e.g., that any $3 \times 3$ block matrix of $\bfG$ is nonsingular. For instance, the $9 \times 9$ matrix formed by picking the first three block columns has determinant $-1$ over $\mathbb{R}$. In the matrix-vector multiplication scenario we can use $\bfG$ as the coding matrix (see Fig. \ref{unidecex}) by setting $m = 9$. This system can tolerate one failure. 

An advantage of this method is that $\bfG$ is binary. Moreover, it has significantly better worst case condition number as compared to the polynomial approach (see Table \ref{table1}, Column 3). However, we are unaware of efficient decoding techniques for these methods. Thus, the decoding complexity is equivalent to Gaussian elimination.
\end{example}




Convolutional codes are another class of erasure codes where messages are encoded into sequences of varying length.
As an example, consider two row vectors in $\mathbb{R}^3$, $\mathbf{u}_0 = [u_{00}~ u_{01}~ u_{02}]$ and $\mathbf{u}_1 = [u_{10}~ u_{11}~  u_{12}]$. These vectors can also be represented as polynomials $\mathbf{u}_i(D) = \sum\limits_{j=0}^{2}u_{ij} D^j$ for $i = 0,1$, where $D$ is an indeterminate. Consider the following encoding of $[\mathbf{u}_0(D) \; \; \mathbf{u}_1(D)]$.
\begin{align*}
[\bfc_0(D) \;\; \bfc_1(D) \;\; \bfc_2(D) \;\; \bfc_3(D)] & = \; \left[\mathbf{u}_0(D) \; \mathbf{u}_1(D) \right] \; \begin{bmatrix}
1 & 0 & 1 & 1 \\
0 & 1 & 1 & D \\
\end{bmatrix}\\
&= \left[\bfu_0(D) \;\; \bfu_1(D) \;\; \left(\bfu_0(D) + \bfu_1(D) \right) \;\; \left(\bfu_0(D) + D\bfu_1(D)\right) \right].
\end{align*}
It is not too hard to see that the polynomials $\mathbf{u}_0(D)$ and $\mathbf{u}_1(D)$ (equivalently the vectors $\mathbf{u}_0, \mathbf{u}_1$) can be recovered from any two entries of the vector $[\bfc_0(D) ~\bfc_1(D)~\bfc_2(D)~\bfc_3(D)]$. For instance, suppose that we only have $\bfc_2(D)$ and $\bfc_3(D)$ where
\begin{align*}
 \bfc_2(D) & =   (u_{00} + u_{10}) + (u_{01} + u_{11})D + (u_{02} + u_{12})D^2, \text{~and} \\
 \bfc_3(D) &=  u_{00}  + (u_{01} + u_{10})D + (u_{02} + u_{11})D^2 + u_{12} D^3.
\end{align*}
Starting with $u_{00}$ from the constant term of $\bfc_3(D)$, one can recover $u_{10}$ from $\bfc_2(D)$ and iteratively $u_{01}$ from $\bfc_3(D)$ and so on. A similar argument applies if we consider a different pair of entries from $[\bfc_0(D) ~\bfc_1(D)~\bfc_2(D)~\bfc_3(D)]$. Distributed matrix-vector multiplication can be embedded into this convolutional code by interpreting the coefficients of the powers of $D$ as the assignments to the workers (see \cite{das2019random, DasR19}).

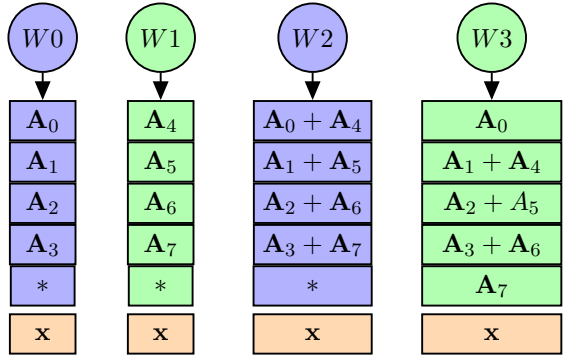
\begin{figure}[t]
\centering
\resizebox{0.5\linewidth}{!}{
\begin{tikzpicture}[auto, thick, node distance=2cm, >=triangle 45]

\draw

    node [sum, fill=blue!30,] (blk1) {$W0$}
    node [sum, fill=green!30,right = 0.7cm of blk1] (blk2) {$W1$}
    node [sum, fill=blue!30,right = 1.2cm of blk2] (blk3) {$W2$}
    node [sum, fill=green!30,right = 1.6cm of blk3] (blk4) {$W3$}

    node [block, fill=blue!30, minimum width = 2.5em, below = 0.4 cm of blk1] (blk11) {$\bfA_{0}$}
    node [block, fill=blue!30,minimum width = 2.5em,below = 0.0005 cm of blk11] (blk12) {$\bfA_{1}$}
    node [block, fill=blue!30,minimum width = 2.5em,below = 0.0005 cm of blk12] (blk13) {$\bfA_{2}$}
    node [block, fill=blue!30, minimum width = 2.5em, below = 0.0005 cm of blk13] (blk14) {$\bfA_{3}$}
    node [block, fill=blue!30,minimum width = 2.5em,below = 0.0005 cm of blk14,minimum width = 2.43em] (blk16) {$*$}
    node [block, fill=orange!30,minimum width = 2.5em,below =  0.1 cm of blk16] (blk17) {$\bfx$}

    node [block, fill=green!30,minimum width = 2.5em,below = 0.4 cm of blk2] (blk21) {$\bfA_{4}$}
    node [block, fill=green!30,minimum width = 2.5em,below = 0.0005 cm of blk21] (blk22) {$\bfA_{5}$}
    node [block, fill=green!30,minimum width = 2.5em,below = 0.0005 cm of blk22] (blk23) {$\bfA_{6}$}
    node [block, fill=green!30,minimum width = 2.5em,below = 0.0005 cm of blk23] (blk24) {$\bfA_{7}$}
	node [block, fill=green!30,minimum width = 2.5em,below = 0.0005 cm of blk24,minimum width = 2.43em] (blk26) {$*$}
	node [block, fill=orange!30,minimum width = 2.5em,below =  0.1 cm of blk26] (blk27) {$\bfx$}
	
    node [block, fill=blue!30,below = 0.4 cm of blk3] (blk31) {$\bfA_0 + \bfA_4$}
    node [block, fill=blue!30,below = 0.0005 cm of blk31] (blk32) {$\bfA_1 + \bfA_5$}
    node [block, fill=blue!30,below = 0.0005 cm of blk32] (blk33) {$\bfA_2 + \bfA_6$}
    node [block, fill=blue!30,below = 0.0005 cm of blk33] (blk34) {$\bfA_3 + \bfA_7$}
    node [block, fill=blue!30,below = 0.0005 cm of blk34, minimum width = 4.5em] (blk36) {$*$}
	node [block, fill=orange!30,minimum width = 4.5em, below =  0.1 cm of blk36] (blk37) {$\bfx$}
	
    node [block, fill=green!30,minimum width = 5.3em, below = 0.4 cm of blk4] (blk41) {$\bfA_{0} $}
    node [block, fill=green!30,minimum width = 5.3em,below = 0.0005 cm of blk41] (blk42) {$\bfA_1 + \bfA_4$}
    node [block, fill=green!30,minimum width = 5.3em,below = 0.0005 cm of blk42] (blk43) {$\bfA_2 + A_5 $}
    node [block, fill=green!30,minimum width = 5.3em,below = 0.0005 cm of blk43] (blk44) {$\bfA_3 + \bfA_6$}
    node [block, fill=green!30,minimum width = 5.3em, below = 0.0005 cm of blk44] (blk46) {$\bfA_7$}
	node [block, fill=orange!30,minimum width = 5.3em, below =  0.1 cm of blk46] (blk47) {$\bfx$}
;
\draw[->](blk1) -- node{} (blk11);
\draw[->](blk2) -- node{} (blk21);
\draw[->](blk3) -- node{} (blk31);
\draw[->](blk4) -- node{} (blk41);

\end{tikzpicture}
}

\caption{\small The scheme corresponding to the approach of \cite{das2019random} as described in Example \ref{eg:matvec}.}
\label{matvecex}
\end{figure}

\begin{example}
\label{eg:matvec}
Consider a system with $N=4$ workers, with  $\gamma_A = \frac{5}{8}$.
We partition $\bfA$ into $m=8$ block-columns of equal size which are denoted as $\bfA_0, \bfA_1, \dots, \bfA_{7}$. So, we have $\calA_0(D) = \bfA^T_{0} + \bfA^T_{1} D +  \bfA^T_2 D^{2} + \bfA^T_3 D^{3}$ and $\calA_1(D) = \bfA^T_{4} + \bfA^T_{5} D +  \bfA^T_6 D^{2} + \bfA^T_7 D^{3}$. The matrices assigned to the $i$-th worker are given by the coefficient of the powers of $D$ in $\bfC_i(D)$, where
\begin{align*}
\left[ \bfC_0(D) \;\;\;  \bfC_1(D)  \;\;\; \bfC_2(D) \;\;\; \bfC_3(D) \right]  \, = \,
 \left[
 \calA_0(D) \;\;\; \calA_1(D) \right] \; \begin{bmatrix}
1 & 0 & 1 & 1 \\
0 & 1 & 1 & D \\
\end{bmatrix} .
\end{align*}
This is illustrated in Fig. \ref{matvecex}. It can be verified that the system is resilient to two failures. Furthermore, it can be shown that the system of equations that the master node has to solve can be put in lower-triangular form upon appropriate permutations. Thus, decoding is quite efficient. We note here that this approach leads to a slightly non-uniform assignment of tasks to the different worker nodes, e.g., $W3$ has one additional matrix-vector product to compute as compared to the other worker nodes. However, this non-uniformity can be made as small as desired by choosing a large enough $m$, while ensuring that the decoding complexity remains low. It also has much better condition number as compared to the polynomial based schemes (see Table \ref{table1}, Column 4). It turns out that multiplying the elements of the encoding matrix by random numbers allows us to provide upper bounds on the worst-case condition number of the recovery matrices (see Table \ref{table1}, Column 5). Decoding in this case requires a least-squares solution; this least-squares solution can be made more efficient by exploiting the sparse nature of the underlying matrices \cite{das2019random}.
%
\end{example}

A fountain coding approach (also known as rateless coding) was presented in the work of \cite{mallickCJ19}. In this scenario, the master node keeps computing random binary linear combinations of the $\bfA_i$'s and sending them to the worker nodes. 
These combinations are chosen from a carefully designed degree sequence. The properties of this degree sequence guarantee with high probability that as long as the receiver obtains $m(1 + \epsilon)$ matrix-vector products where $\epsilon>0$ is a small constant, the receiver can decode the desired result (the result is asymptotic in $m$). Furthermore, this decoding can be performed using a so-called peeling decoder, which is much simpler than running full-blown Gaussian elimination. In a peeling decoder, at each time instant, the receiver can find one equation where there is only one unknown. This is important as in the large $m$ regime, the cubic complexity of Gaussian elimination would be unacceptably high, whereas the peeling decoder has a complexity $\approx m \log m$.





\subsection{Distributed Matrix-Matrix Multiplication}


The situation is somewhat more involved when consider the distributed computation of $\bfA^T\bfB$. In this case one needs to consider the joint design of the coded versions of the blocks of $\bfA$ and $\bfB$ ({\it cf.} eq. (\ref{eq:block_decomp_A_B})). This topic was the focus of the so-called algorithm-based fault-tolerance (ABFT) techniques \cite{huangA84}\cite{jouA86} in the 80's. However, ABFT techniques result in sub-optimal recovery thresholds. The work of \cite{yu2017polynomial}, presented an elegant solution to this problem based on polynomials which matches a corresponding lower bound on the threshold in certain cases. Interestingly, work on embedding matrix-matrix multiplication into the structure of polynomials was considered much earlier in the work of \cite{yagle1995fast}; however, this was in the context of speeding up the computation rather than straggler resilience.

The basic idea(s) of using polynomials for matrix-matrix multiplication have already been illustrated by Examples \ref{eg:polycode} and \ref{eg:entangled_poly_code} in Section \ref{sec:problem_form}. In what follows we present a more in-depth discussion of these techniques along with a host of other approaches that have been considered in the literature. The first idea along these lines in \cite{yu2017polynomial} corresponds to the case of $p=1$ and arbitrary $m$ and $n$ (using the notation introduced in Section \ref{sec:problem_form}). As before, polynomial $\bfA(z) = \sum_{i=0}^{m-1} \bfA_i z^i$. However, the second polynomial with coefficients $\bfB_j, j = 0, \dots, n-1$ needs to be chosen more carefully. The underlying simple and useful trick is to choose $\bfB(z)$ in such a way that $\bfA_i^T \bfB_j$ for $i = 0, \dots, m-1, j = 0, \dots, n-1$ appear as coefficients of $z^l$ for $l = 0, \dots, mn-1$ of the polynomial $\bfA^T(z) \bfB(z)$. Reference \cite{yu2017polynomial} proposes to use
\begin{align*}
\bfA(z) \; &= \; \sum_{j=0}^{m-1} \bfA_j z^j \; \; \; \; \textrm{and} \; \; \; \; \bfB(z) \; = \; \sum_{j=0}^{n-1} \bfB_j z^{jm},\\
\textrm{so that } \; \bfA^T(z) \bfB(z) \; & = \; \sum_{j=0}^{m-1} \sum_{k=0}^{n-1} \bfA^T_j \bfB_k z^{j + km}.
\end{align*}
The $i$-th worker node is assigned $\bfA(z_i)$ and $\bfB(z_i)$ so that the storage fractions are $\gamma_A = 1/m$ and $\gamma_B = 1/n$. It is tasked with computing $\bfA^T(z_i) \bfB(z_i)$.
Evidently, $\bfA^T(z) \bfB(z)$ can be interpolated to determine the intended result as long as the master node obtains $mn$ distinct evaluations of it. This solution is such that the computational load and the communication load on each worker is $1/mn$-th of the original. It also achieves the optimal recovery threshold (under communication load limitations on the worker nodes). Furthermore, the decoding complexity corresponds to running $\frac{rw}{mn}$ polynomial interpolations of a degree-$mn-1$ polynomial. Nevertheless, this technique has serious numerical stability issues stemming from the ill-conditioned nature of the Vandermonde structured recovery matrices discussed before ({\it cf. } Section \ref{sec:distr_mat_vec}). 

A generalization of this approach for matrix-matrix multiplication when $p > 1$ was considered in \cite{yu2018straggler} and \cite{duttaFHJCG19} around the same time. This was earlier examined in the matrix-vector context when each worker only gets subvectors of $\bfx$ in the work of \cite{dutta2016short}. The work in \cite{dutta2016short} can be considered as a special case of this result when $n=1$. However, the threshold in \cite{yu2018straggler} is better than \cite{dutta2016short}. Our discussion below, loosely follows the presentation in \cite{yu2018straggler}. Note that when $p=1$, our unknowns are precisely $\bfA_i^T \bfB_j, i =0, \dots, m-1, j = 0,\dots,n-1$. However, when $p=2$ (for instance), the unknowns now involve the sum of certain terms. Indeed, when $m=n=p=2$, we have
\begin{align*}
\bfA^T \bfB &= \begin{bmatrix}
\bfA^T_{00} \bfB_{00} + \bfA^T_{10} \bfB_{10} & \bfA^T_{00} \bfB_{01} + \bfA^T_{10} \bfB_{11}\\
\bfA^T_{01} \bfB_{00} + \bfA^T_{11} \bfB_{10} & \bfA^T_{01} \bfB_{01} + \bfA^T_{11} \bfB_{11}
\end{bmatrix}.
\end{align*}
Recall, that our goal is to form polynomials $\bfA(z)$ and $\bfB(z)$ with coefficients from $\bfA_{ij}, i = 0,\dots, m-1, j = 0, \dots,  p-1$ and $\bfB_{kl}, k=0, \dots, p-1, l = 0, \dots, n-1$ such that the useful terms appear as appropriate coefficients of consecutive powers of $z$ in $\bfA^T(z) \bfB(z)$. When $p > 1$ (unlike $p=1$), the presence of interference terms becomes unavoidable. Nevertheless, one can choose $\bfA(z)$ and $\bfB(z)$ in such a way that we can interpolate the useful terms along with interference terms at the master node. This can lead to a strictly better threshold as indicated in Example \ref{eg:entangled_poly_code}. We refer the reader to the full details in \cite{yu2018straggler}. For $m=n=p=2$, we choose
\begin{align*}
\bfA(z) &= \bfA_{00} + \bfA_{10}z + \bfA_{01}z^{2} + \bfA_{11} z^{3},\\
\bfB(z) &= \bfB_{10} + \bfB_{00}z +  \bfB_{11} z^{4} + \bfB_{01}z^{5}, \text{~so that}\\
\bfA^T(z)\bfB(z) &= (*) + (\bfA^T_{00} \bfB_{00} + \bfA^T_{10} \bfB_{10})z + (*) z^2 +  (\bfA^T_{01} \bfB_{00} + \bfA^T_{11} \bfB_{10}) z^3 \\
&+ (*) z^4 + (\bfA^T_{00} \bfB_{01} + \bfA^T_{10} \bfB_{11}) z^5 + (*) z^6 + (\bfA^T_{01} \bfB_{01} + \bfA^T_{11} \bfB_{11}) z^7,
\end{align*}
where $(*)$ in the expression above refers to an interference term that we are not interested in. It can be observed that $\bfA^T(z)\bfB(z)$ is a matrix polynomial of degree-7 and can therefore be interpolated as long as eight distinct evaluations are obtained. In general, the result of \cite{yu2018straggler} shows that the threshold of their scheme is $\tau = pmn + p-1$. The scheme can be decoded efficiently via polynomial interpolation. However, the numerical stability issue in this case is even more acute as the degree of the fitted polynomial is $pmn+p-2$, i.e., much higher. 

Recently, there have been some contributions in the literature that attempt to address the numerical stability issues associated with polynomial based approaches. In \cite{das2019random}, the authors demonstrate that convolutional codes can be used for matrix multiplication as well. They also demonstrate a computable upper bound on the worst case condition number of the recovery matrices. This approach allows for schemes that are significantly better in terms of the numerical stability. The authors in \cite{FahimC19}, propose an alternate approach where the underlying polynomial scheme now operates in the basis of orthogonal polynomials such as Chebyshev polynomials. They show that the condition number of the recovery matrices can be upper bounded polynomially in the system parameters (as long as the number of stragglers is a constant), unlike real Vandermonde matrices where the condition number grows exponentially. The recent work of \cite{ramamoorthyT19} presents a different approach wherein polynomials are evaluated at structured matrices such as circulant permutation and rotation matrices. The worst case condition numbers obtained by this scheme are much lower as compared to \cite{FahimC19} (assuming that the number of stragglers is a constant).


\begin{example}
We now present an experimental comparison of the polynomial code and the convolutional code approach for computing $\bfA^T \bfB$ with $r=w=9000$ and different $t$ (see Table \ref{table:time}). We set up a cluster in the Amazon Web Services (AWS) cloud with {\it one} {\tt t2.2xlarge} machine as the master node and $N = 11$ {\tt t2.small} worker nodes. We considered a system with $p=1,m=n=3$ so that the threshold $\tau =9$. The entries in Table \ref{table:time} correspond to the worst case computation time of each worker node for different values of $t$. We picked the set of workers that correspond to the worst condition number for both schemes. For both methods, it can be seen that while the worker computation time increases roughly linearly with $t$, the decoding time does not change. We note here the computational load on the worker nodes in the convolutional code approach is slightly higher than the polynomial code approach. This difference can be made as small as desired with higher subpacketization \cite{das2019random}. Note however that the condition number of the convolutional code is multiple orders of magnitude smaller. Our code implements the to and from communication from the master node to the workers sequentially; parallel implementations can further reduce these values.

\begin{table}[t]
\centering
\caption{{\small Comparison of polynomial code and convolutional code method in terms of worker computation time, total communication time (in parentheses), decoding time and condition number.}}
\begin{tabular}{c c c c c c}
\hline
\multirow{2}{3 cm}{$\; \; \; \; \; \; \; \; \;$ Methods} & \multicolumn{3}{c}{Worker Comp. and Comm. Time (in $s$)} & $\; \; \; $Decoding Time  & \multirow{2}{2.8 cm}{$\;$ Condition Number} \\
\cline{2-4}
& $t = 12k $ & $\; \; \;  t = 18k $ & $t = 24k $ & $\; \; $ (in seconds)& \\ \hline
Polynomial Codes \cite{yu2017polynomial} & $6.8  \;(5.0)$ & $10.8  \;(6.9)$ & $14.2  \;(8.6)$ & $\; \; \; $ $2.9 \sim 3.0$ & $24753.93$  \\ \hline
Convolutional Codes \cite{das2019random} & $7.9  \;(6.7)$ & $12.2  \;(8.9)$ & $15.7  \;(11.9)$ &  $\; \; \; $  $4.8 \sim 4.9$ & $152.12$ \\ \hline
\end{tabular}%
\label{table:time}%
\vspace{-0.2in}
\end{table}%

\end{example}





\section{Opportunities for future work}

\label{sec:opps_future_work}
The discussion in the preceding sections has hopefully convinced the reader that the area of coded matrix computation is a growing one and that there is ample scope to contribute towards it in various ways. We now outline some outstanding issues that require closer attention from the research community as a whole.

The vast majority of work in this area has considered distributed schemes for computing $\bfA^T \bfB$ for arbitrary matrices $\bfA$ and $\bfB$. However, in several practical scenarios, these matrices are sparse. This can change the computational complexity calculation significantly. We illustrate this by considering matrix-vector multiplication. 
If $\bfA$ (of dimension $t \times r$) is such that each column contains at most $s$ non-zero entries then computing $\bfA^T \bfx$ takes $\approx 2rs$ flops. Suppose that we apply the polynomial solution of Section \ref{sec:distr_mat_vec}. In this situation, each coded matrix $\tilde{\bfA}_l$ has approximately $sm$ non-zero entries per column in the worst case (assuming $sm < t$). The worker node that computes $\tilde{\bfA}_l^T \bfx$ will therefore require $(1/m) \times 2rsm = 2rs$ flops. This means that in the worst case each worker node has the ``same" computational load as computing $\bfA^T \bfx$, i.e., the computational advantage of distributing the computation may be lost. Table \ref{table2} tabulates the time for computing $\tilde{\bfA}^T_l \bfx$ (for a $30,000 \times 30,000$ $\bfA$) using the solution of \cite{lee2018speeding} for a system with $N=15$ worker nodes with a threshold of $\tau=m=12$, for two kinds of sparse matrices: (i) a $\bfA$ that has the $\beta$-diagonal structure where the diagonal and $\beta$ off-diagonal terms are non-zero, and (ii) a $\bfA$ where the non-zero entries are chosen at random. Table \ref{table2} also lists the time of computing an uncoded matrix vector product, i.e., $\bfA_i^T\bfx$. It is clear the worker node computation time increases significantly for the coded case. We note here that this is an issue with other papers \cite{yu2017polynomial}\cite{yu2018straggler}\cite{das2019random}\cite{DasR19}\cite{RamTV19} as well. The fountain coding approach for the matrix-vector case \cite{mallickCJ19} fares better here because with high probability the linear combination generated by the master node has low weight. However, \cite{mallickCJ19} does not provide provable guarantees on the recovery threshold and does require rather high values of $m$. This was also considered in \cite{wang2018coded} for the matrix-matrix case, though it is unclear whether their scheme respects the storage constraints on the workers as formulated in Section \ref{sec:problem_form}. The recent work of \cite{wangLSY19} makes progress on this problem. Reference \cite{wangLSY19} defines the ``computational load'' of a given coding solution in the matrix-vector case as the number of non-zeros elements of the corresponding coding matrix. It contains a discussion about lower bounds and achievability schemes for this metric.

\begin{table}[t]
\centering
\caption{{\small Comparison of worker computation times when $\bfA$ is a sparse matrix. The first row lists the worker time for finding $\bfA_l^T \bfx$, while the second lists the time to find $\tilde{\bfA}^T_l \bfx$.
In each column, the number in parentheses is for the case when $\bfA$ has a $\beta$-diagonal structure and the other number is for a sparse random $\bfA$.}}
\begin{tabular}{c c c c}
\hline
Percentage of zero-entries & $90\%$ & $80\%$ & $70\%$ \\ \hline
Time for uncoded case in $ms$ & $11.9 (13.3)$ & $22.1 (22.7)$ & $36.8 (35.4)$  \\ \hline
Time for coded case in $ms$ & $109 (83.8)$ & $110.1 (104.3)$ & $122.2 (108.2)$ \\ \hline
\end{tabular}%
\label{table2}%
\vspace{-0.2in}
\end{table}%

Throughout this review article, we have highlighted the role of embedding an erasure code into a distributed matrix computation problem. As we have shown, in the computation context special attention needs to be paid to the numerical stability of the recovery of $\bfA^T \bfB$. Much of existing work does not provide guarantees on the worst-case or average-case condition numbers and this is an important direction that needs to be pursued. There have been some initial results in this area \cite{ramamoorthyT19}\cite{das2019random}\cite{FahimC19}, but much remains to be done.

The majority of existing work only deals with the recovery threshold ({\it cf.} Section \ref{sec:problem_form}) which is in one-to-one correspondence with treating an erasure as a failed node. However, recovery threshold(II) considers a more fine-grained model where different worker nodes operate at different speeds. The systematic design of schemes that provably leverage partial computations by the worker nodes is interesting. Reference \cite{RamTV19}, considers the case of matrix-vector multiplication, but systematic extensions to the matrix-matrix multiplication case would be of interest.
\section{Conclusion}
\label{sec:conclusion}
We surveyed the state of the art schemes of distributed matrix-vector and matrix-matrix multiplication in this article. MATLAB and Python code for several of the schemes in this paper can be downloaded from \cite{das_github}. These problems are of significant interest as several basic machine learning algorithms use them repeatedly in various intermediate steps. We have also pointed out various avenues for future work.



\end{document}